# A QM/MM molecular dynamic simulation and vibrational spectroscopic study of 6-azaCytidine and cytidine

Marawan Ahmed[*], Anoja P. Wickrama Arachchilage and Feng Wang

eChemistry Laboratory, Faculty of Life and Social Sciences, Swinburne University of Technology, Melbourne, Victoria 3122 Australia.

**Abstract**

The conformational and vibrational properties of 6-azaCytidine (6-azaC) and Cytidine (Cyt) have been studied in gas phase and aqueous environment. The two most stable conformers of each molecule have been identified and investigated in more details. The stability order of the two most stable conformers of both molecules is changing going from gas phase to solution. The first solvation shell of both molecules has been studied using QM/MM-MD (Quantum Mechanics/Molecular Mechanics-Molecular Dynamic) simulations. The Raman vibrational spectra in gas phase and aqueous solution have been simulated and compared with the available experimental data. The two molecules exhibit certain differences reflected in their spectra as a result of structural and conformational differences.

[*]To whom correspondence should be addressed. E-mail: mmahmed@swin.edu.au, Tel: 61-3-9214 8785.

# 1- INTRODUCTION

Studying the molecular properties at the atomic level of details in different solution environments is one of the hottest topics in computational chemistry. In real world, most of chemical and/or biological reactions are taking place not in gas phase/isolated environment but in solution environment. Typically, more than 60% of the human body is composed of water, which makes water (aqueous environment) to be the solvent of choice for dealing with biological systems [1]. For laboratory experiments, it's possible to carry out certain experiment at different solvent environments of different polarities. The choice of certain solvent to carry out a particular experiment depends on several factors such as the solubility of the substances, solvent effect on the substances and several more. For these reasons, it's not strange that modelling solution environment occupies a very important place in computational chemistry [2].

Cytidine nucleoside analogues represent a very important class of antiviral agents. They may interfere with certain nuclear enzymes, acting as enzyme inhibitors, or be incorporated in a growing nucleic acid chain causing cell death [3]. Owing to the high rate of mutations especially in viruses, the need of new antiviral drugs is increasingly important as well as insight understanding of the molecular behaviour of the current known agents [4, 5]. As an integrated part of this insight understanding comes the modelling of such important molecules in aqueous environment where they really exist. A very important antiviral agent is the 6-azaCytidine nucleoside analogue where the $6^{th}$ position of the base ring of Cytidine is replaced by a nitrogen atom. 6-azaC is a very active nucleoside analogue with antifungal, antiviral and antitumor activities [6-9].

The X-ray crystal structures of Cytidine identified the presence of the anti-conformation in solid state [10]. This is in accordance with the majority of other pyrimidine nucleosides where the anti-conformation has been shown to be the prevalent conformational form in solid state and/or aqueous solution [11-13]. In an interesting study by Mishchuk et al. and applying the semiempirical MNDO/H method, it has been shown some of the most stable structures of 6-azaC belongs to both the syn and anti-conformations [14]. As experimental evidences identify the anti-conformation to be the prevalent conformational structures for the majority of Cytidine nucleosides, our main focus will be directed against this conformation. In that case, only the orientation of the exocyclic arm ($C_{5'}$-$O_{5'}$) will now be the major criteria for determining the overall 3D structure of a given nucleoside. The orientation of this exocyclic arm can be defined by the γ angle ($\angle \mathbf{C_{(3')}}$-$\mathbf{C_{(4')}}$-$\mathbf{C_{(5')}}$-$\mathbf{O_{(5')}}$ ($^\mathbf{O}$)), and according to this angle, each nucleoside can be denoted to be *gauche* (<90°) or *trans* (>90°). This is of course together with the puckering and pseudorotation considerations of the sugar moiety, which are known to influence both the geometry as well as the biological activity [15-17].

Cytidine nucleosides analogues have been the main focus of lots of both experimental and theoretical vibrational (IR/Raman) as well as NMR studies [9, 15, 18-21]. As a result of the sensitivity of both IR and Raman vibrational spectroscopy to conformational and/or tautomeric states changes, vibrational spectroscopy is an integrated tool for studying the fine molecular structure of nucleosides [22]. Particularly the formation of inter/intra molecular H-bonds are very well studied using vibrational spectroscopy [23-26]. Given the fact that most of these biomolecules exert their functions in aqueous environment, it's very important to study the effect of such aqueous media in different properties of these molecules, including the effect on the vibrational spectra [9, 19, 27]. It's also widely known that the biological activities of these biomolecules are greatly linked with their fine molecular structure [17, 28, 29]. Grouping together these pieces of information, we conclude that modelling of these Cytidine nucleosides in aqueous environment is inevitable to build a solid knowledge about the molecular structure of these molecules.

Modelling of aqueous environment can be achieved utilizing either implicit and/or explicit solvation models. In the implicit solvation models, solvent is expressed as a featureless medium and the polarity is expressed by the dielectric constant (ε) [30]. The higher the dielectric constant (ε), the higher the polarity. These kinds of models are usually of lower computational cost than the explicit ones and are often used in QM calculations [30]. The implicit models ignore direct interaction of the solute atoms with water molecules. The PCM model by Tomasi and co-workers [31-33] is one of the most used models. In explicit solvation models, solvents molecules are modelled explicitly in the calculations and this makes the calculation more computationally expensive. This is inevitable when it's essential to model solute-solvent interaction explicitly. For explicit modelling of water molecules, the TIP3P [34] model and its congeners, for example, the TIP4P, TIP5P and the SPC/Fw…etc., are the often used models in biological simulations [35, 36].

In the current study, we will try to (i) understand the basic differences between the conformational structures of 6-azaC and Cyt in gas phase and in aqueous solution, (ii) studying the first hydration shell of both molecules through interaction with explicit water molecules using explicit solvent QM/MM-MD simulations and (iii) correlating these differences by comparing both simulation and available experimental results of vibrational spectra.

## 2- METHODS AND COMPUTATIONAL DETAILS

Chemical structures and atomic numbering of 6-azaCytidine (**6-azaC**, **I**) and Cytidine (**Cyt, II**) are presented in Figure 1. As indicated previously [37-40], rotations of the glycosyl $\chi$ angle and the $\gamma$ angle, combined with the sugar puckering produce a number of nucleoside conformers. The most stable conformation of nucleosides **I** and **II** are identified using a fully relaxed two-dimensional (2D) potential energy surface scan, i.e., $V(\chi,\gamma)$, based on the density functional theory (DFT) based B3LYP/6-31G model, where the glycosyl $\chi$ angle and the $\gamma$ angle are defined as $\chi=\angle O_{(4')}-C_{(1')}-N_{(1)}-C_{(2)}$ and $\gamma=\angle C_{(3')}-C_{(4')}-C_{(5')}-O_{(5')}$, respectively.

The obtained energy minimum structures of **I** and **II** on the potential energy surfaces are further optimized using the B3LYP/6-311++G** model. No imaginary frequencies are found for both nucleosides in the harmonic frequency calculations, indicating that the energy minimum structures obtained are true minima rather than saddle points. Sugar puckering [41-44] of the nucleosides, which is determined by the pseudorotational phase angle (P°) and the amplitude of puckering ($\nu_{max}$), is obtained using the PROSIT tool [39].

The IR/Raman vibrational spectra are calculated using the B3LYP/6-311++G** model in gas phase and aqueous solution ($\varepsilon= 78.36$) applying the PCM model [45] and using the most stable conformer of each molecule. All calculations were performed using Gaussian 09 (G09) computational chemistry package [46].

For hydration shells calculation, the optimized nucleoside was immersed in a cubic box of water with a buffer region from the boundary of about 20 Å. QM/MM dynamics is performed using the dispersion corrected DFTB Hamiltonian for the QM system which is composed of the nucleoside only and using the AMBER suite [47, 48]. The MM subsystem was treated using the flexible SPC/Fw water model as implemented in AMBER and applying the ff99SB force field [36, 49]. The system was first minimized then heated to 300 °K in the NVT ensemble for 2 ps then relaxed classically for 2 ns in the NPT ensemble. Again the system is relaxed in the NPT ensemble but applying the dispersion corrected DFTB (Density Functional Tight Binding) Hamiltonian in a QM/MM-MD routine for the nucleoside only subsystem and for 1 ns [50-54]. The DFTB Hamiltonian is known to give excellent structural results for nucleosides and similar biomolecules [55-57]. The system was again relaxed in the NPT ensemble for 2 ps applying the PBE/6-31G* [58] Hamiltonian for the nucleoside only in a QM/MM-MD routine and using the newly developed Gaussian-AMBER interface [48, 59]. The PBE Hamiltonian is known to give an excellent description for H-bonding [60, 61]. The QM/MM-MD run was then extended by 20 ps production simulation. In all stages, periodic boundary conditions are applied and a small time step of 0.5 fs was utilized to integrate the equations of motion. Electrostatic interactions were truncated after 15 Å and for long range electrostatic interactions, particle mesh Ewald (PME) [62] was utilized.

## 3. RESULTS AND DISCUSSION

### 3.1. Geometrical properties

Figure 2 represents the 3D structure of the gas phase most energetically favourable conformational structures of both **I** and **II.** Potential sugar-base intramolecular H-bonds are also shown as dashed lines "spider web". Corresponding geometrical data are presented in Table 1. Studying the geometrical structure of nucleosides and their analogues is of great importance owing to the influence on the biological activity as shown before, but this can be considered as a challenging task [15-17]. Usually, ground state properties are determined based on thermodynamical equilibrium between different interchanging conformers. The numerical value for a given property is the ensemble average of the numerical values of this property for each conformer weighed by the probability of finding this conformer at a given temperature. Even the simplest nucleoside has a myriad of conformational structures making the decision of choosing which conformer to use for data interpretation be a very difficult decision [7, 15, 21, 63]. In most cases, the most energetically favourable structure(s) is the conformer of choice. However, if the energy differences between low energy conformers are too low, i.e., at the few kJ.mol$^{-1}$ scale, situation is getting complicated, and usually MD simulation is the best solution.

Table 1 reports the optimized geometrical and energetic parameters of the two most stable conformers both in gas phase and water together with the available crystal structure data. Crystal structure data were extracted directly from the Crystal Information Files (.cif) we got from the Cambridge Structure Database [64]. As we can see in the table, the two most stable conformers of both molecules are belonging to the *anti* category of conformations. For **I**, the most noticeable difference between the crystal structure and the optimized geometrical structures is the orientation of the base regarding to the sugar moiety, this orientation is defined by the $\chi$ angle. In the crystal structure, this angle is given by -78.74° which lies within the *syn* definition of pyrimidine nucleosides. In the optimized structures and in all cases, the values range between (-160° to -170°) which impose the pure *anti* definition of these nucleosides. This is not unexpected as pointed out previously giving the fact that in crystal structure other intermolecular and crystal packing forces may intervene [65]. This intervention can favour conformation that are not usually populated either in gas phase or solution were such intermolecular interactions are minimal. In contrast to **I**, both conformers of **II** show a better agreement with the crystal structure geometrical data with a pure anti orientation, the $\chi$ angle range is given by (-160° to -170°) for both the optimized and crystal data structure. The major discrepancy between gas phase and solution for **I** is the change in the Pseudorotation angle ($P°$) which increases from 184° in gas phase to 205.87° in solution for **IA**, and from 186.92° to 208.62° for **IB**.

Regarding the orientation of the exocyclic ($C_{5'}$-$O_{5'}$) arm which is defined by the $\gamma$ angle ($\angle C_{(3')}$-$C_{(4')}$-$C_{(5')}$-$O_{(5')}$ ($^O$)), it's found that the gas phase most energetically favourable conformations are those belonging to the *gauche* category of conformations. This conformation is strongly stabilized in gas phase by the aid of several intramolecular H-bonds. For example, in **IA** the $O_{5'}$-H⋯$O_{4'}$ is given by 2.509 Å and the $O_{5'}$-H⋯$N_6$ intramolecular H-bond is given by 2.127 Å. The $O_{5'}$-H⋯$N_6$ H-bonds is slightly longer in **IB** and is given 2.431 Å. This is as a result of the change of the conformational structures of **I** from the gauche conformation in **IA** to the trans conformation in **IB**. For **II**, the **IIA** *gauche* conformation is stabilized by a strong $C_6$-H⋯$O_{4'}$ H-bond. On the other hand, the **IIB** *trans* conformer is stabilized by two H-bonds, the $C_6$-H⋯$O_{4'}$ H-bond which is given by 2.216 Å and the O5'-H⋯$O_{4'}$ bond which is given by 2.404 Å.

Table 2 reports the total and free energy differences and relative population ratios (%) of the compounds under study at the B3LYP/6-311++G** in gas phase and solution. Population ratio was obtained through differences in the Gibbs free energy (**ΔG**) which takes into account both the entropy and the effect of temperature. Free energy calculations were carried out at the experimental temperature (298 °K). In each case, the most stable conformer from each pair was assumed to be unity. Final population ratio was obtained by normalization and summation again to unity. The following equation has been applied:

$$\frac{n_j}{n_i} = e^{-\Delta G/k_b T}$$

In the above equation, $n_j$ is the proportion of the least stable conformer in each pair, $n_i$ is taken as 1, ΔG is free energy difference, $k_b$ (8.31 J K$^{-1}$ mol$^{-1}$) is the gas constant and T is the experimental room temperature (298 °K) [9, 21]. These ratios are just for demonstration as actual ratios should take into account all potential populated conformers at the specified temperature.

Regarding the order of stability and as we can see in Table 2, it's found that **IA** is more stable than **IB** both in gas phase (ΔT.E.= 4.783 kj.mol$^{-1}$) and in solution (ΔT.E.= 1.355 kj.mol$^{-1}$). However, at the experimental temperature and in water, **IB** is more stable (**ΔG**=1.027 kj.mol$^{-1}$). In contrast to **I** which shows a noticeable energy change, **II** doesn't show much energy change and both conformers are almost degenerate (ΔT.E.= 0.197 kj.mol$^{-1}$) in gas phase. Moreover, **IIB** is of slightly higher stability in water (ΔT.E.= 1.289 kj.mol$^{-1}$). Considering the change in free energy (**ΔG**), it's found that the most stable conformer in gas phase is different from the one found in solution for both **I** and **II**. Interestingly and in both molecules, the *trans* conformer is of higher stability in water while the *gauche* conformer is more stable in gas phase. This higher stability in water and at the experimental temperature guarantees higher population of the *trans* conformers under the specified experimental conditions. These population values are given by 60% for **I** and by 68% for **II**. This conformational preference (particularly for Cytidine **II**) of **IIB** over **IIA** in solution is also reflected in their vibrational analysis in aqueous solution where the *trans* conformer (**IIB**) shows more agreement to experimental spectra than the *gauche* (**IIA**) conformer.

## 3.2. Molecular dynamics and nucleoside solvation shell/shells.

In most cases, the first hydration shell of a given organic molecule in water is determined by the number of H-bonds donors or acceptors (heteroatoms) in this molecule. This can be considered as a reasonable approximation given the fact that strong H-bonds with explicit water molecules are usually taking place with heteroatoms (N, O, S …etc.) rather than carbon atoms [66-68]. The second and most difficult challenge is to arrange such waters around that organic molecule, i.e. how waters in the first hydration shell are oriented and posed [69-71]. Previous studies on similar systems have selected the most stable hydrated clusters based on the energy of these clusters [69-71]. Given the fact that there could be a huge number of acceptable low energy arrangements of such water molecules, those properties that take into account statistical averaging of these arrangements are the most acceptable. Typical example is the Radial pair Distribution Function (RDF) which can be calculated and extracted directly from MD trajectories. RDF also takes into account the exchange and diffusion motions of water molecules in the hydration shell/shells.

Figures 3 (a-c) displays the RDF plots for some important nucleoside atoms with regards to the surrounding water molecules. Three atomic sites have been chosen, the $O_2$-$H_w$, the $(O_{5'})H_{5'}$-$O_w$ and the $(C_{5'})H_{5'}$-$O_w$. As we can see in the figure, heteroatom mediated H-bonds results in a well-defined hydration shells characterized by a sharp peak at the first and second hydration shells. For example, for both molecules, the RDF of the $O_2$-$H_w$ and the $(O_{5'})H_{5'}$-$O_w$ distance has two sharp peaks at 1.6 Å and 3.35 Å. This indicates a very well structured hydration shells. However, for the $(C_{5'})H_{5'}$-$O_w$ distance, the first solvation shell peak has a maximum value above 2.5 Å, also, the peak is flat and not well defined. This is not unexpected given the fact that carbon mediated H-bonds are belonging to the very weak category of H-bonds.

Figure 4 displays selected snapshots of the MD trajectories for both molecules at a 5 ps interval (5ps, 10ps, 15ps and 20 ps) including waters within 4 Å of each nucleoside. The $O_2$ atom is always di- or tri-coordinated to the surrounding waters ($H_w$). Other heteroatoms are always mono- or di-coordinated. The overall hydration shells distributions for both molecules are very similar. This is a direct consequence for the similar overall 3D structures of both molecules. We were also interested in tracking the dihedral angles change during the simulation time. We chose to trace the two most important dihedrals, $\chi°$ ($\angle O_{(4')}$-$C_{(1')}$-$N_{(1)}$-$C_{(2)}$) and $\gamma°$ ($\angle C_{(3')}$-$C_{(4')}$-$C_{(5')}$-$O_{(5')}$). As we can see in Figure 5(a) and in most of the simulation time, both molecules adopt an *anti*-conformation for the $\chi°$ angle (below -90° line) that defines the orientation of the base moiety with regard to the sugar moiety. For the $\gamma°$ (Figure 5b) angle, both molecules adopt a pure *trans*-conformation all over the whole simulation time period.

### 3.3. Raman spectra

In the vibrational data analysis section, only the *trans-gauche* conformers of 6-azaC(**IB**) and Cyt (**IIB**) will be considered given the fact that they are more stable in aqueous phase than their *gauche-gauche* counterparts. Figure 6 displays the comparative experimental and simulated Raman spectra of the most stable conformer of each molecule, 6-azaC (**IB**) and Cyt(**IIB**) in aqueous phase at the 500-1800 cm$^{-1}$ region. Supporting info files contain also the results of both conformers against experiment to show this preference (S1). We have also included the results of the IR spectra in the same context and it also shows better agreement for the Cyt(**IIB**) conformer that the Cyt(**IIA**) conformer with the IR experimental spectra in solution (S2). Vibrational wavenumbers have been calculated at the B3LYP/6-311++G** and owing to the good agreement with the available experimental spectra ( < 2000 cm$^{-1}$), we didn't shift or scale the simulated spectra. Particularly for Cyt, the **IIB** conformer exhibits a better agreement with the available experimental data which is consistent with the slightly higher energy barrier than that for 6-azaC between the two most populated conformers for each molecule. The available experimental Raman spectra of both molecules exhibit a reasonable agreement with the simulated spectra, particularly in the region 500 – 1800 cm$^{-1}$. For example, the experimental Raman spectra of 6-azaC and Cyt exhibit prominent peaks at 1200-1400 cm$^{-1}$ and ~1500-1700 cm$^{-1}$, which can be observed in the simulated Raman spectra. Table 1 presents the assignments of prominent frequencies of 6-azaC and Cyt Raman spectra in water phase along with the available experimental frequencies. The experimental Raman spectrum of Cyt exhibits a peak at 1205 cm$^{-1}$ for $C_{(6)}$-H and $C_{(5)}$-H out of phase bending vibration [21], which is observed at 1210 cm$^{-1}$ of the simulated spectrum of Cyt. The available experimental Raman spectral assignment of 6-azaC at water does not provide lot of information. The experimental Raman spectrum of 6-azaC exhibits a prominent peak at 758 cm$^{-1}$ for pyrimidine ring vibrational modes [9]. The simulated Raman spectrum of 6-azaC exhibits a peak at 768 cm$^{-1}$ for this particular vibrational mode, providing an excellent agreement between theory and experiment.

The fingerprint regions of the Raman spectra of two molecules are different. For example, 6-azaC exhibits a prominent single peak at ~1300 cm$^{-1}$ for C-H bend vibrational mode in pyrimidine ring and the sugar moiety. However, at the same region, Cyt exhibits a prominent split peaks for C-H bend frequencies of pyrimidine ring (~1250 cm-1) and sugar moiety (~1300 cm-1). The aqueous phase most stable conformer of Cyt (**IIB**) exhibits a strong intra-molecular H-bond interaction at H$_{(6)}$···O$_{(4')}$ (Figure 2d), which may be the reason for separating pyrimidine ring and sugar C-H bending frequencies. The Raman frequencies at the region of 1400 – 1800 cm$^{-1}$ of 6-aza are scattered with number of prominent peaks, with respect to that of Cyt. The frequencies of this particular region is attributed by the C=N stretch vibrational modes. The C$_{(4)}$=N$_{(3)}$ stretch is common in both the molecules, which is shown by an intense peak at ca. 1540 cm$^{-1}$. It is seen that the substitution at C$_{(6)}$ site in Cyt by nitrogen in 6-aza does not effect on the C$_{(4)}$=N$_{(3)}$ stretch frequency. In the Raman spectrum of 6-azaC, an extra peak at 1626 cm$^{-1}$ represents the C$_{(5)}$=N$_{(6)}$ stretch, which shifts the C$_{(2)}$=O$_{(2)}$ stretch frequency of 6-azaC by 62 cm$^{-1}$ towards higher frequency from that of Cyt.

Figure 7 presents the second cluster of Raman vibrations of 6-azaC and Cyt in water phase at the region 2800-4000 cm$^{-1}$. Note that the experimental spectra of two compounds are available for the region of 400-2000 cm$^{-1}$, due to the congested and overlapping peaks at the 2000 – 4000 cm$^{-1}$ region resulted by the number of molecular interactions in the water phase [9]. However, we simulated the complete Raman spectra of these two molecules, to study the stretching frequencies, which demonstrate the intramolecular interactions. The substitution at C$_{(6)}$ site of Cyt by a nitrogen alters the Raman spectra of 6-azaC significantly from Cyt counterpart. For example, in 6-azaC the C-H and N-H/O-H stretch frequency regions are clearly separated with ca. 400 cm$^{-1}$ of frequency gap. However, the gap between C-H and N-H/O/H stretch frequencies in Cyt is smaller (ca. 200 cm$^{-1}$) compared to 6-azaC, due to the blue shifting of hexagon C-H stretch frequencies and red shifting of O(2')-H stretch frequency of Cyt with respect to 6-azaC frequencies. According to the Figure 2, the bond distance of O$_{(2')}$-H···O$_{(2)}$ is 1.772 Å and 1.972 Å, in Cyt(**IIB**) and 6-azaC(**IB**), respectively. The O$_{(2')}$-H···O$_{(2)}$ H-bond interaction is stronger in Cyt(**IIB**) than that of 6-azaC(**IB**), which may be the reason for 185 cm$^{-1}$ of red shifting of O$_{(2')}$-H stretch in Cyt from 6-azaC counterpart. The stretch vibrations of $NH_2$ group in both molecules produce peaks at the same frequency. However, O-H stretch frequencies of both molecules exhibit significant frequency shifts. For example, $O_{(5')}$-H stretch is at 3815 cm$^{-1}$ in Cyt, whereas it is at 3705 cm$^{-1}$ in 6-azaC. The $N_{(6)}$ in 6-azaC facilitates a H-bond interaction of $O_{(5')}$-$H_{(5')}\cdots N_{(6)}$, in which the bond distance is 2.431 Å. This H-bond interaction may be the reason for blue shifting in $O_{(5')}$-H stretch frequency of 6-azaC by 110 cm$^{-1}$ from Cyt.

**CONCLUSIONS**

The geometrical structures of two nucleosides 6-azaCytidine and Cytidine have been compared in gas phase and aqueous solution. The *gauche-gauche* conformer of both molecules seems to be the most stable in gas phase whereas the *trans-gauche* conformers are more stable in water. The first hydration shell of both molecules are very similar as well as the RDFs around some selected atomic sites as obtained through QM/MM-MD simulations. The experimental Raman spectra of these two molecules provided the information about the molecular structural differences. The simulated Raman spectra of two molecules are significantly different, particularly at the region of 2800 – 4000 cm$^{-1}$. The substitution at $C_{(6)}$ of Cytidine by a nitrogen in 6-azaCytidine significantly alters the Raman spectral properties of both molecules, which reflects their molecular structural changes and the intra-molecular interactions.

**ACKNOWLEDGEMENTS**

MA and AP acknowledge the Swinburne University Postgraduate Research Award (SUPRA). We thank the Victorian Partnership for Advanced Computing (VPAC) and Swinburne University supercomputing (Green/Gstar) for the support on the computing facilities. The National Computational Infrastructure (NCI) at the Australian National University and the Victorian Life Sciences Computation Initiative (VLSCI) on its Peak Computing Facility at the University of Melbourne (an initiative of the Victorian Government, Australia) under the Merit Allocation Scheme (MAS) are acknowledged.

**Table 1:** Selected gas phase/aqueous solution geometric parameters of the nucleoside derivatives 6-azaCytidine (**I**) and Cytidine (**II**) in comparison with the available experimental results.

| Parameters | 6-azaCytidine (**I**) | | | | | Cytidine (**II**) | | | | |
|---|---|---|---|---|---|---|---|---|---|---|
| | Conf.1 (**IA**) | | Conf.2 (**IB**) | | Expt.[c] | Conf.1 (**IIA**) | | Conf.2 (**IIB**) | | Expt.[c] |
| | Vacuum[a] | Water[b] ($\varepsilon = 78.36$) | Vacuum[a] | Water[b] ($\varepsilon = 78.36$) | | Vacuum[a] | Water[b] ($\varepsilon = 78.36$) | Vacuum[a] | Water[b] ($\varepsilon = 78.36$) | |
| $R_5$ (Å) | 7.470 | 7.471 | 7.467 | 7.470 | 7.456 | 7.477 | 7.476 | 7.494 | 7.486 | 7.39 |
| $R_6$ (Å) | 8.163 | 8.162 | 8.160 | 8.159 | 8.168 | 8.244 | 8.239 | 8.248 | 8.241 | 8.14 |
| $\chi = \angle O_{(4')}\text{-}C_{(1')}\text{-}N_{(1)}\text{-}C_{(2)}$ (°) | -162.65 (*anti*) | -161.744(*anti*) | -170.36(*anti*) | -169.210(*anti*) | -78.74(*syn*) | -170.27 (*anti*) | -170.47 (*anti*) | -171.63 (*anti*) | -169.161 (*anti*) | -162.85 (*anti*) |
| $\gamma = \angle C_{(3')}\text{-}C_{(4')}\text{-}C_{(5')}\text{-}O_{(5')}$ (°) | 61.07(*gauche*) | 60.418(*gauche*) | 165.427(*trans*) | 163.548(*trans*) | 55.71(*gauche*) | 56.41(*gauche*) | 54.97(*gauche*) | 178.438(*tr*) | -179.53(*tr*) | 47.13(*gauche*) |
| $\omega = \angle O_{(4')}\text{-}C_{(4')}\text{-}C_{(5')}\text{-}O_{(5')}$ (°) | -58.40(*gauche*) | -59.16(*gauche*) | 47.98(*gauche*) | 46.322(*gauche*) | -61.14(*gauche*) | -63.91(*gauche*) | -65.03(*gauche*) | 59.67(*gauche*) | 61.37(*gauche*) | -70.37(*gauche*) |
| $P$ (°) | 184.01 | 186.92 | 205.87 | 208.62 | 15.12 | 168.83 | 177.51 | 148.74 | 153.24 | 9.15 |
| $\nu_{max}$ | 29.43 | 30.68 | 33.43 | 35.07 | 37.33 | 30.90 | 32.19 | 35.15 | 35.12 | 37.88 |
| $<R^2>$ (a.u.) | 4041.658 | 4023.132 | 4073.568 | 4039.806 | | 4219.048 | 4176.93 | 4495.944 | 4486.692 | |
| $\mu$ (D) | 6.944 | 9.858 | 6.535 | 9.339 | | 6.067 | 9.169 | 6.376 | 9.570 | |
| Total Energy ($E_h$) | -907.443716 | -907.468508 | -907.441894 | -907.467992 | | -891.428646 | -891.453817 | -891.428571 | -891.454308 | |
| ZPE (kcal.mol$^{-1}$) | 141.681 | 141.610 | 141.384 | 141.337 | | 148.856 | 148.670 | 148.888 | 148.667 | |
| T.E. +ZPE ($E_h$) | -907.217932 | -907.242838 | -907.216584 | -907.242758 | | -891.191429 | -891.216896 | -891.191302 | -891.217391 | |
| Sugar type | C3'-exo/anti | C3'-exo/anti | C3'-exo/anti | C3'-exo/anti | C3'-endo | C2'-endo/anti | C2'-endo/anti | C2'-endo/anti | C2'-endo/anti | C3′-endo/anti |

[a] B3LYP/6-311++G**.
[b] B3LYP/6-311++G** level of theory and applying the PCM model.
[c] Data are extracted directly from the .cif files that we got from the Cambridge Structural Database.

**Table 2**: Relative energies and populations (%) of the different conformers of the studied compound in gas phase and solution, at the B3LYP/6-311++G** level of theory.

|  | 6-azaCytidine (**I**)/Vacuum | | 6-azaCytidine (**I**)/Water | | Cytidine (**II**)/Vacuum | | Cytidine (**II**)/ Water | |
|---|---|---|---|---|---|---|---|---|
|  | Conf.1 (**IA**) | Conf.2 (**IB**) | Conf.1 (**IA**) | Conf.2 (**IB**) | Conf.1 (**IIA**) | Conf.2 (**IIB**) | Conf.1 (**IIA**) | Conf.2 (**IIB**) |
| $\Delta(T.E.)$/kJ.mol$^{-1}$ | 0 | 4.784 | 0 | 1.355 | 0 | 0.197 | 1.289 | 0 |
| $\Delta(T.E.+ Z.P.E)$/kJ.mol$^{-1}$ | 0 | 3.539 | 0 | 0.210 | 0 | 0.333 | 1.300 | 0 |
| $\Delta G^{298}$/kJ mol$^{-1}$ | 0 | 2.245 | 1.027 | 0 | 0 | 0.189 | 1.877 | 0 |
| $P$/%$^a$ | 71% | 29% | 40% | 60% | 52% | 48% | 32% | 68% |

$^a$ The most stable conformer from each pair was assumed to be unity. Final population ratio was obtained by normalization and summation again to unity.

**Table 3:** Comparison between selected simulated and experimental Raman vibrational frequencies and their assignments of the most stable conformer in water for 6-azaCytidine (**IB**) and Cytidine (**IIB**).

| 6azaCytidine (**IB**) | | Assign. | Cytidine (**IIB**) | | Assign. |
|---|---|---|---|---|---|
| $\nu$ (cm$^{-1}$) | | | $\nu$ (cm$^{-1}$) | | |
| Sim.[a] | Exp.[b] | | Sim.[a] | Exp.[c] | |
| 3745.79 | | $\nu O_{(3')}$-H | 3815.30 | | $\nu O_{(5')}$-H |
| 3711.41 | | $\nu N_{(4)}$-H-H$_{asym}$ | 3721.33 | | $\nu N_{(4)}$-H-H$_{asym}$ |
| 3705.66 | | $\nu O_{(5')}$-H | 3717.32 | | $\nu O_{(3')}$-H |
| 3648.46 | | $\nu O_{(2')}$-H | 3595.36 | | $\nu N_{(4)}$-H-H$_{sym}$ |
| 3586.98 | | $\nu N_{(4)}$-H-H$_{sym}$ | 3463.13 | | $\nu O_{(2')}$-H |
| 3200.79 | | $\nu C_{(5)}$-H | 3256.82 | | $\nu C_{(6)}$-H |
| 3105.83 – 3004.93 | | $\nu$C-H (sugar ring) | 3223.81 | | $\nu C_{(5)}$-H |
| 1669.74 | | $\nu C_{(2)}=O_{(2)}$ | 3097.56 – 3016.50 | | $\nu$C-H (sugar ring) |
| 1626.23 | | $\nu C_{(5)}=N_{(6)}$ | 1662.12 | 1669 | $C_{(6)}$-H/$C_{(5)}$-H bend + $\nu C_{(5)}=C_{(6)}$ |
| 1624.23 | 1612 | $N_{(4)}$-H-H scissoring | 1648.74 | 1658 | $N_{(4)}$-H-H scissoring |
| 1535.46 | | $\nu C_{(4)}=N_{(3)}$ | 1607.57 | 1655 | $\nu C_{(2)}=O_{(2)}$ |
| 1483.74 | | $\nu C_{(4)}$-$N_{(4)}$ | 1544.64 | 1528 | $\nu C_{(4)}=N_{(3)}$ |
| 1292.55 | | C-H out of plane bend | 1267.93 | 1230 | C-H bend (sugar ring) |
| 768.12 | 758 | Ring vibrations | 1210.44 | 1205 | $C_{(6)}$-H/$C_{(5)}$-H bend |

[a] at the B3LYP/6-311++G** level of theory and applying the PCM model.
[b] Raman spectra in water, see [22].
[c] Raman spectra in water, see [24].

**Figure 1**: The 2D structures (a) 6-azaCytidine **I** and (b) Cytidine **II**.

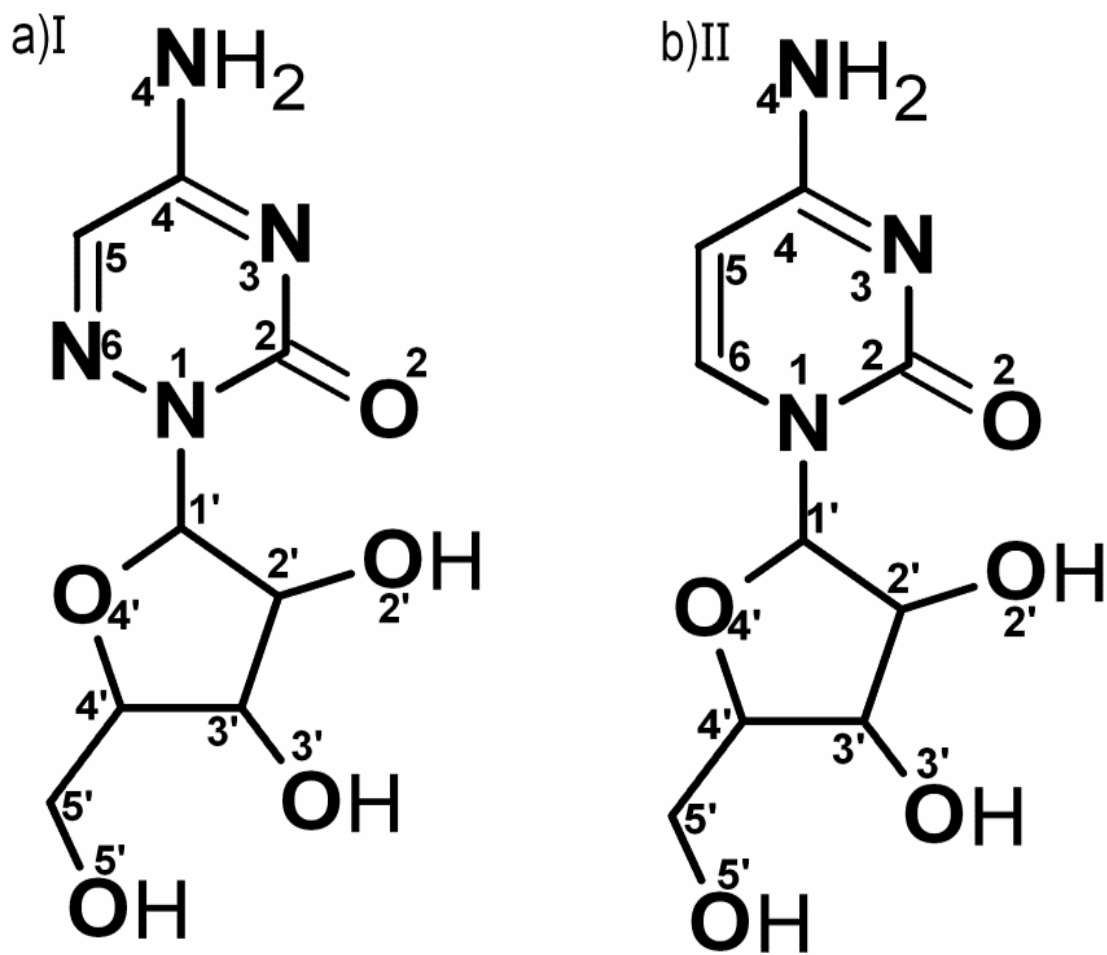

**Figure 2**: The 3D structures of the two most stable conformers of each of the molecules under study (a) 6-azaCytidine **IA,** (b) 6-azaCytidine **IB,** (c) Cytidine **IIA** and (d) Cytidine **IIB.**

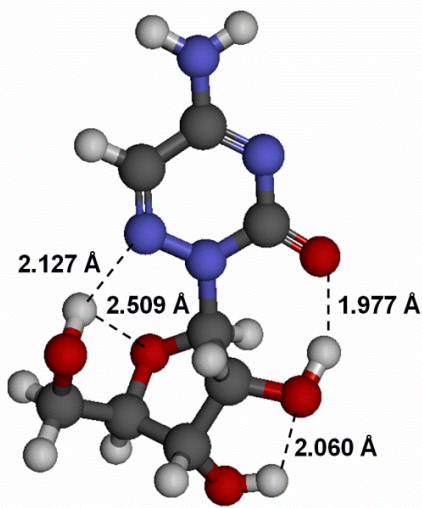
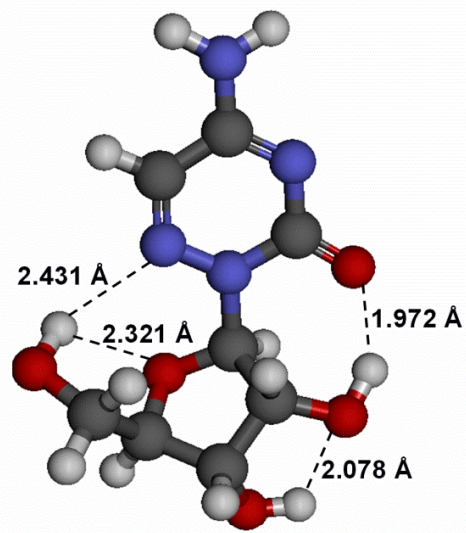
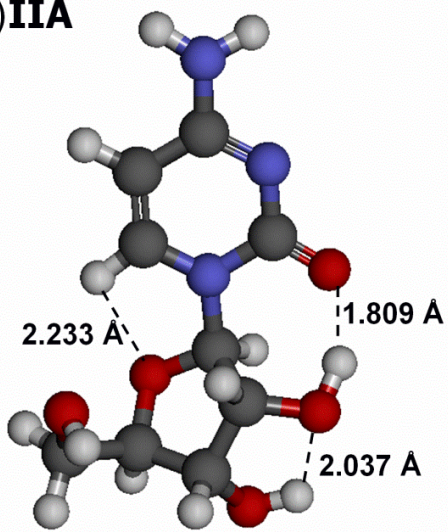
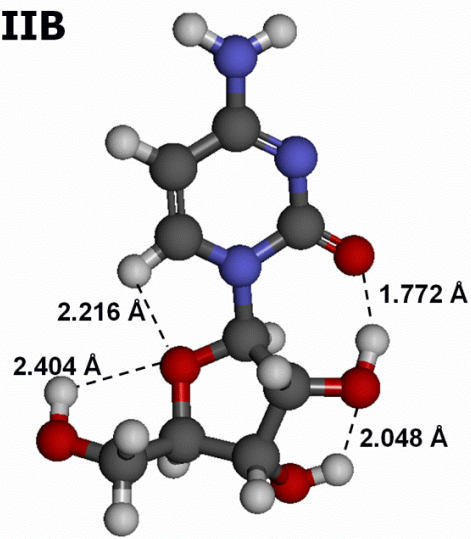

**Figure 3**: Radial distribution functions (RDFs) for distances between selected nucleosides atoms and the surrounding water molecules, (a) $O_2$-$H_w$, (b) $(O_{5'})H_{5'}$-$O_w$ and (c) $(C_{5'})H_{5'}$-$O_w$.

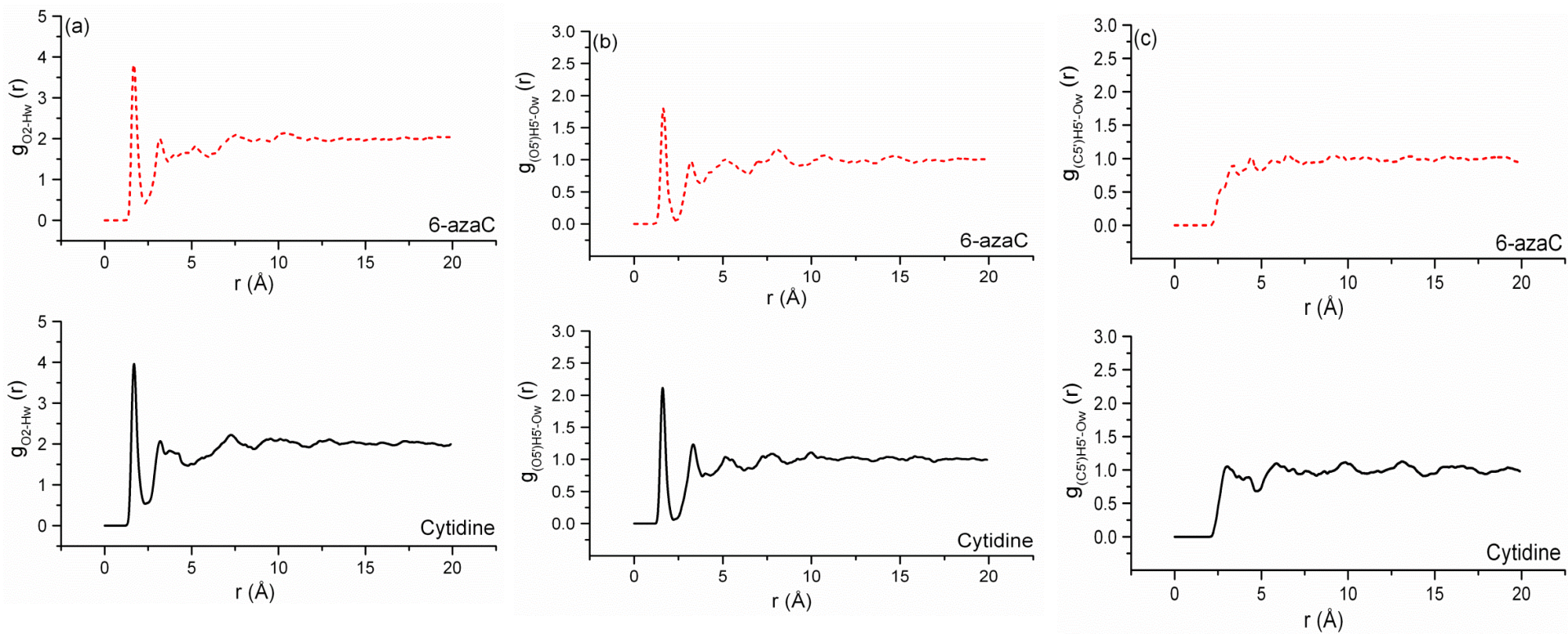

**Figure 4:** Selected trajectory snapshots at a 5 ps time interval (5 ps, 10 ps, 15 ps and 20 ps) of 6azaCytidine **I** (a1-a4) and Cytidine **II** (b1-b4).

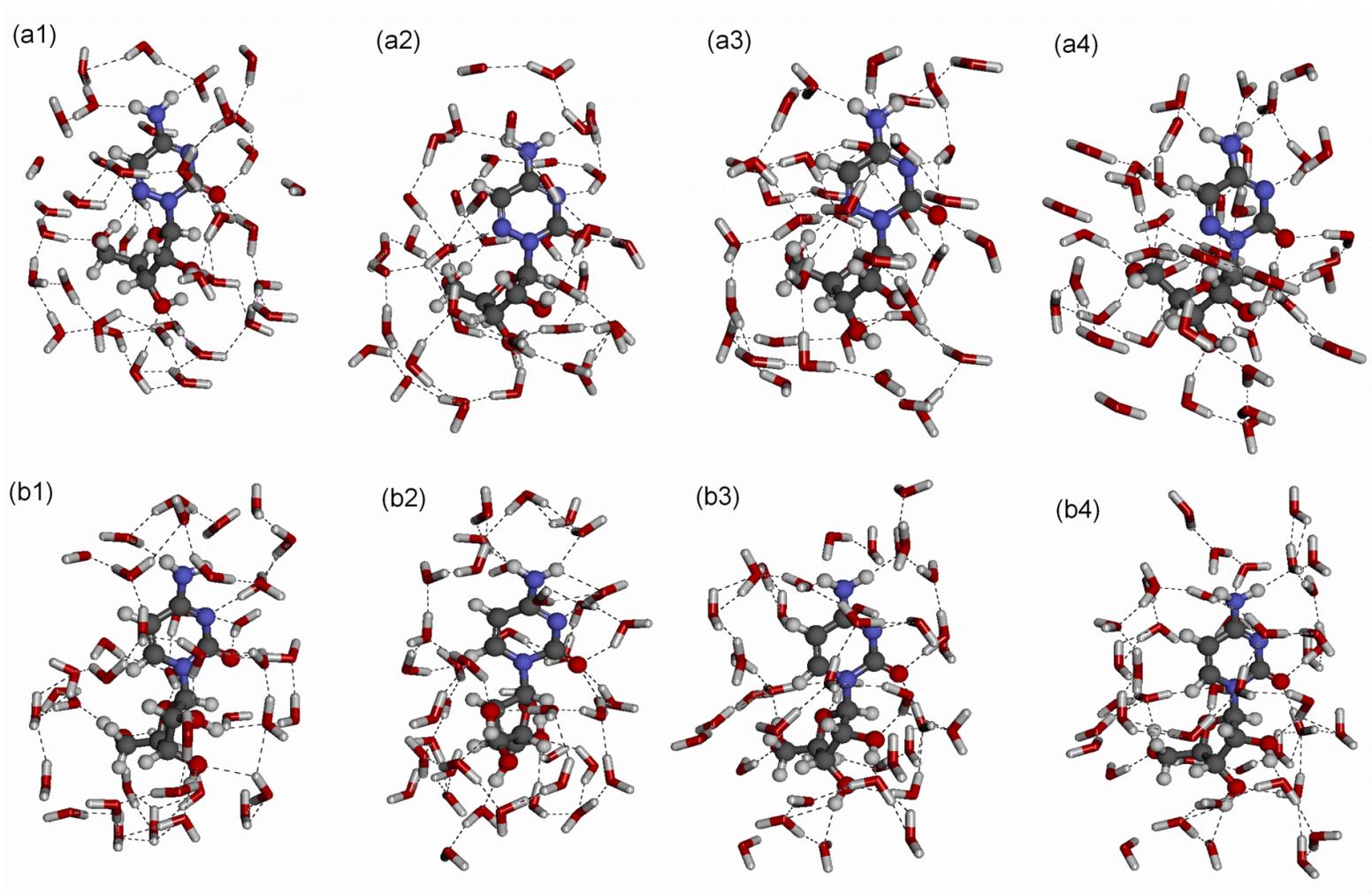

**Figure 5**: Time evolution of selected two dihedrals (a) $\chi°$ ($\angle O_{(4')}-C_{(1')}-N_{(1)}-C_{(2)}$) and (b) $\gamma°$ ($\angle C_{(3')}-C_{(4')}-C_{(5')}-O_{(5')}$) during the 20 ps simulation time.

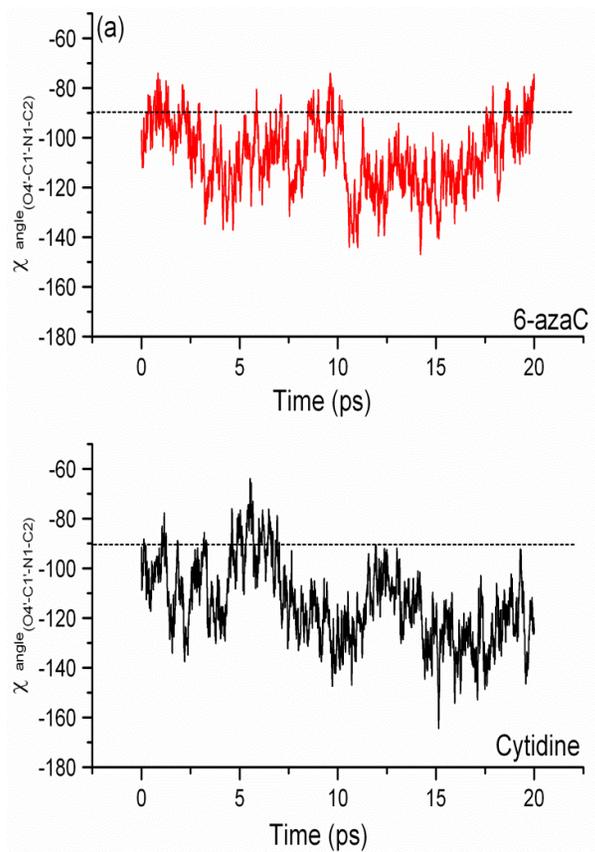
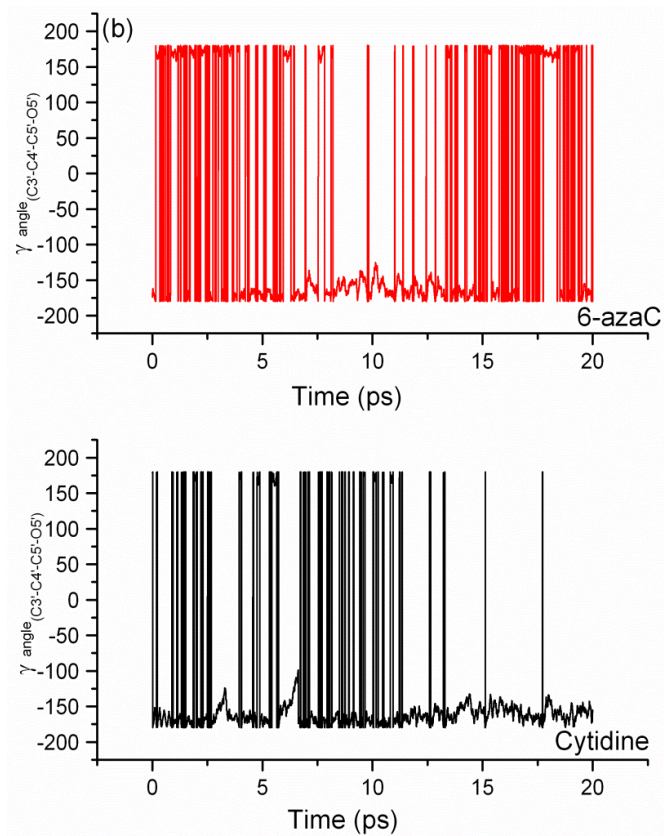

**Figure 6:** Comparative Raman spectra of 6-azaCytidine (**IB**) and Cytidine (**IIB**) in water with the experimental spectra at the region 500-1800 cm$^{-1}$.

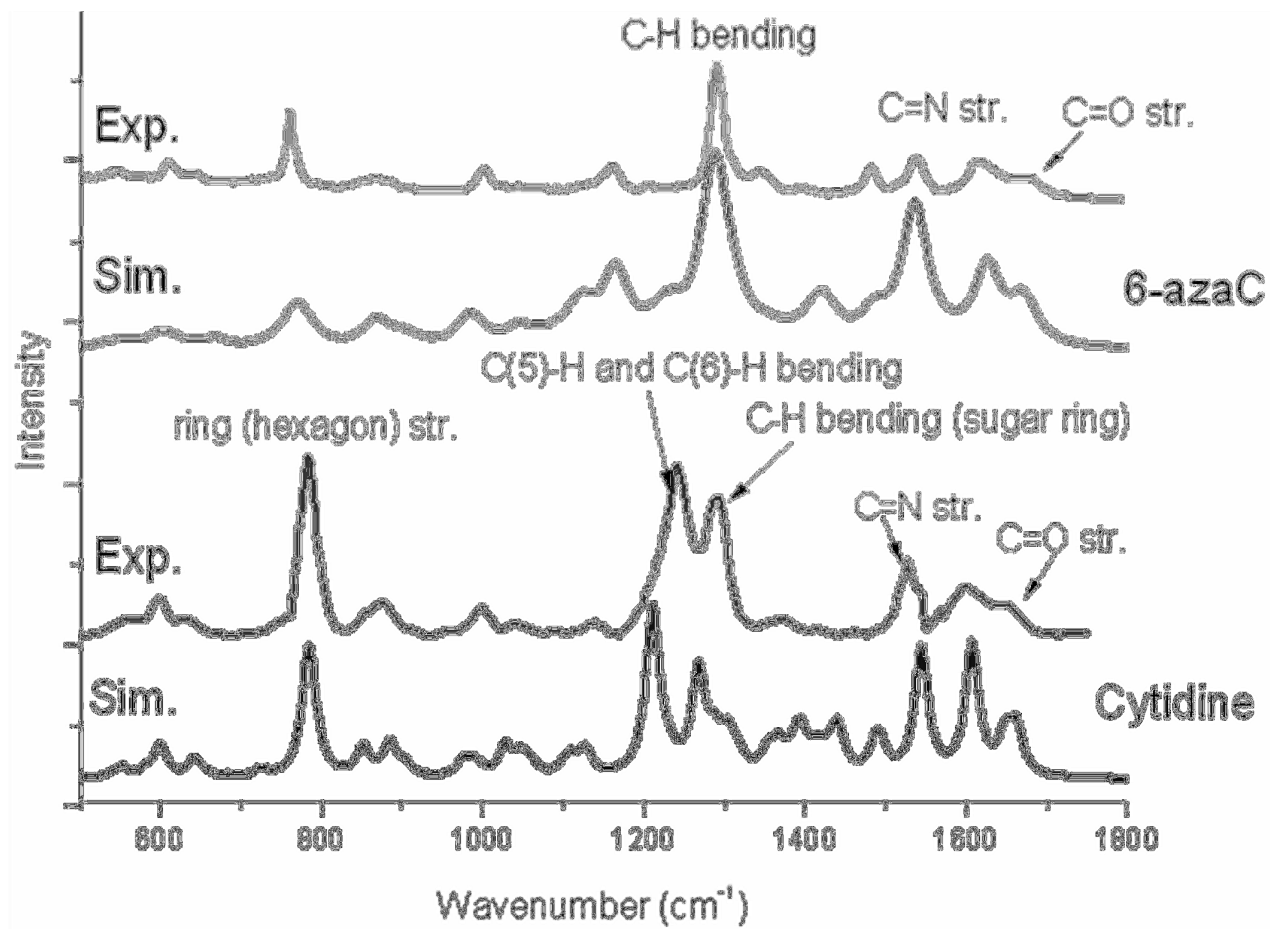

**Figure 7:** Comparative Raman spectra of 6-azaCytidine (**IB**) and Cytidine (**IIB**) in water at the region 2800-4000 cm$^{-1}$.

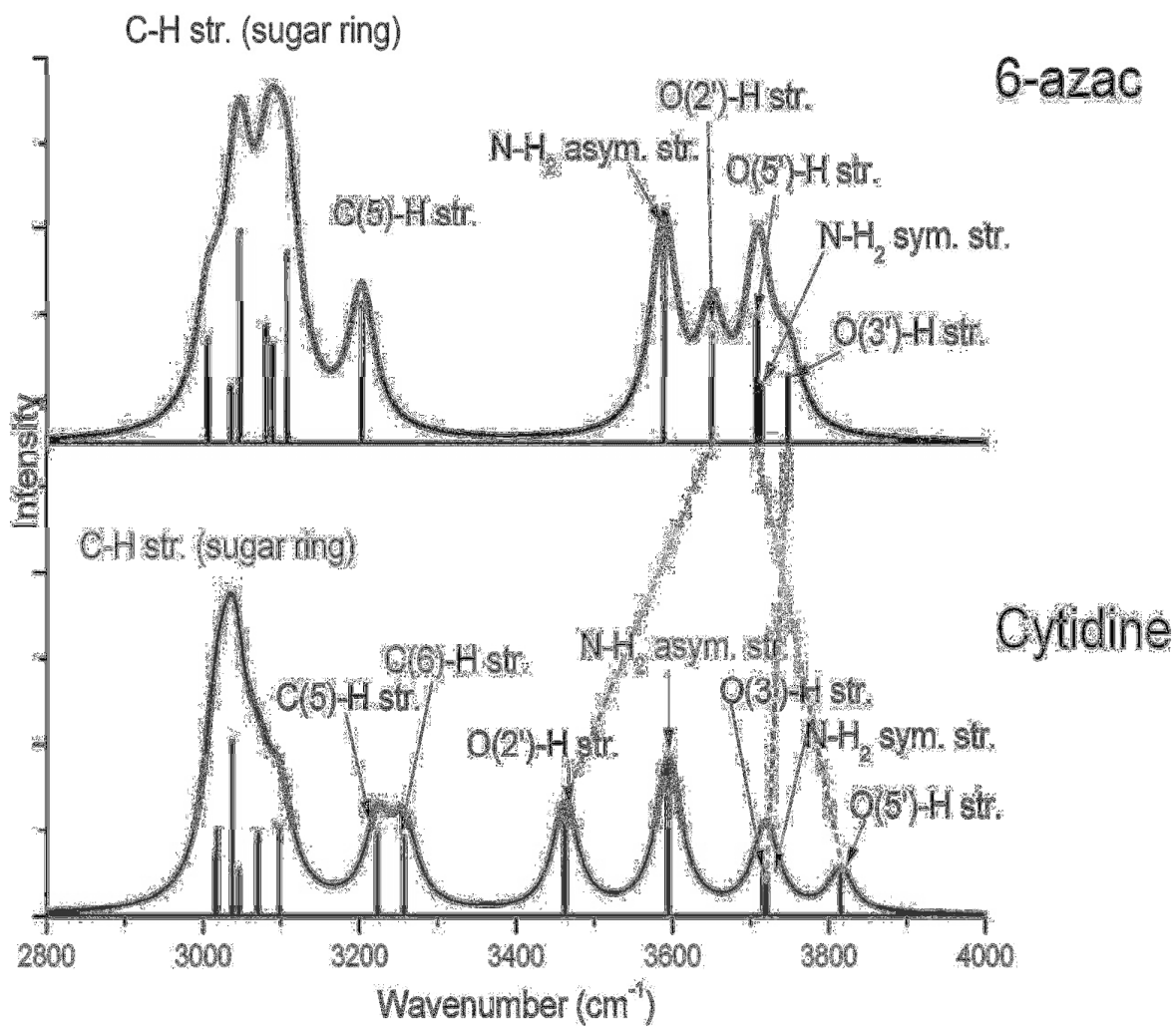

**S1**: Comparative **Raman** spectra of the Cytidine conformers (**IIA/IIB**) in water at the region 200-2000 cm$^{-1}$.

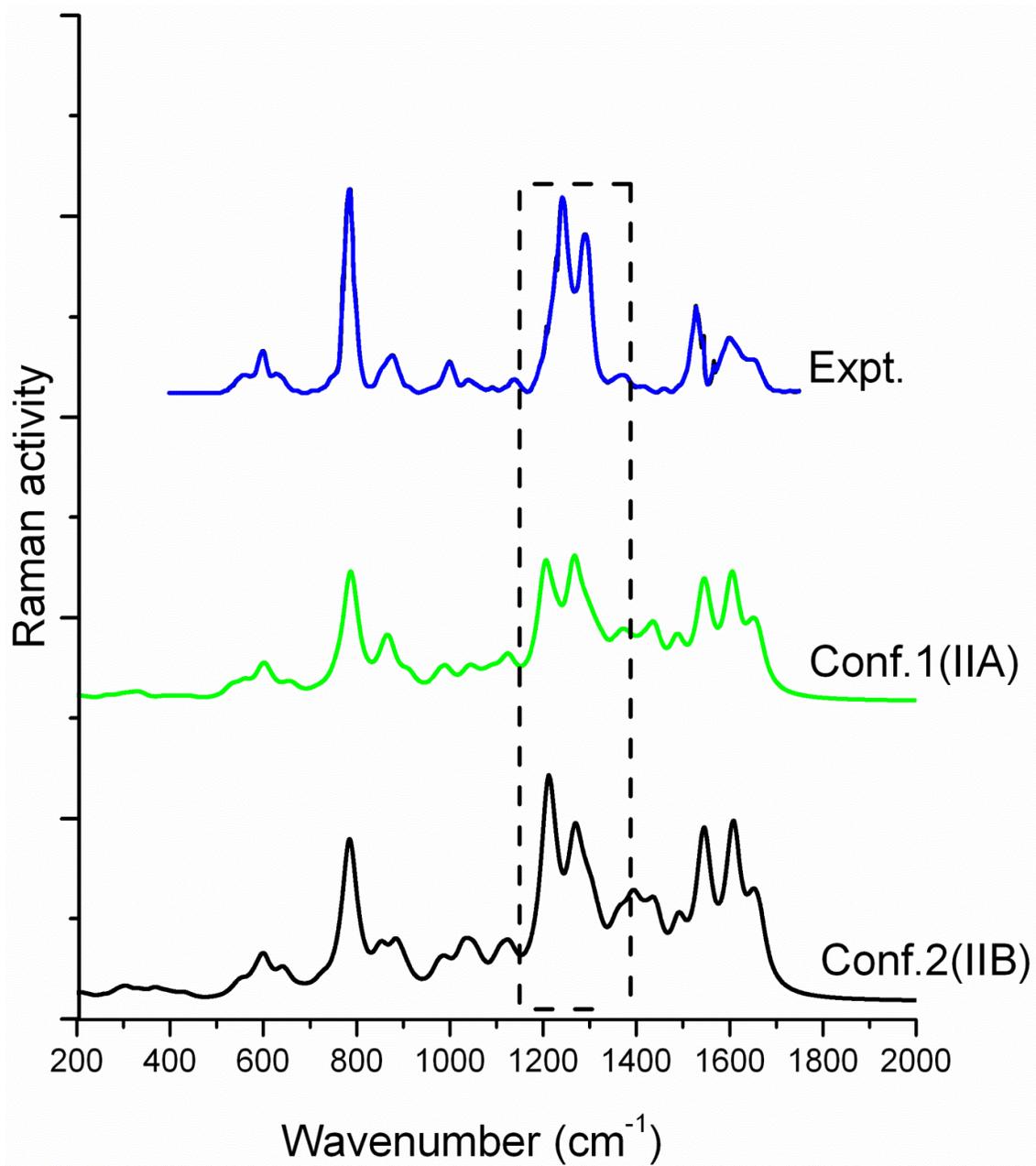

Cytidine_Raman_Theory_Expt._Water
FWHM = 20
B3LYP/6-311++G** /Unscaled

**S2**: Comparative **IR** spectra of the Cytidine conformers (**IIA/IIB**) in water at the region 800-1900 cm$^{-1}$.

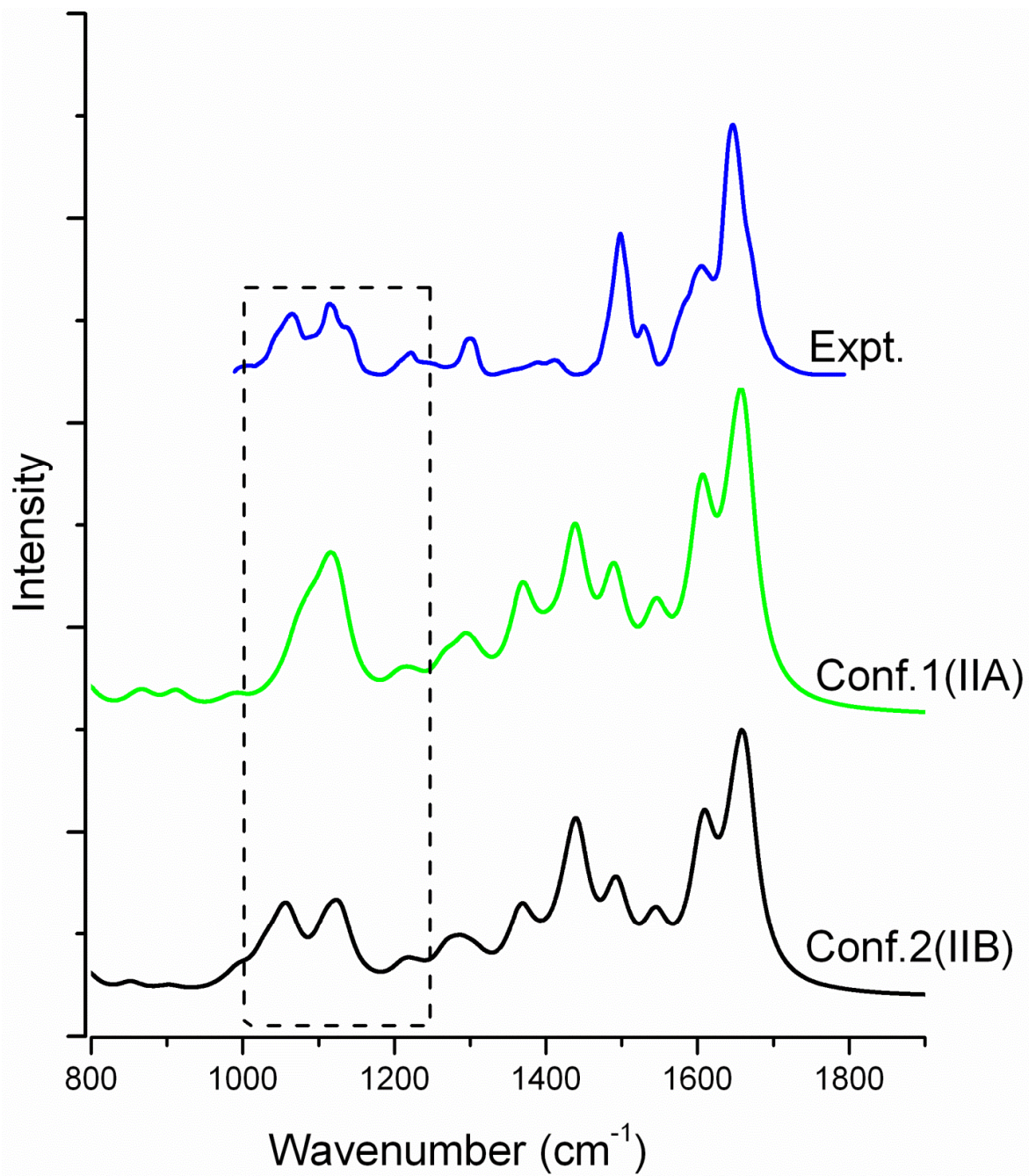

Cytidine_INFRARED_Theory_Expt._Water
FWHM = 20
B3LYP/6-311++G**/Unscaled